\documentclass[fleqn,usenatbib]{mnras}

\usepackage{newtxtext,newtxmath}

\usepackage[T1]{fontenc}

\DeclareRobustCommand{\VAN}[3]{#2}
\let\VANthebibliography\thebibliography
\def\thebibliography{\DeclareRobustCommand{\VAN}[3]{##3}\VANthebibliography}

\usepackage{graphicx}	
\usepackage{amsmath}	
\usepackage{multirow} 

\title[Wind and Jet features]{Signatures of winds and jets in the environment of supermassive black holes}
\author[B. Bandyopadhyay et al.]{
Bidisha~Bandyopadhyay$^{1}$\thanks{E-mail: bidisharia@gmail.com (BB)},
Christian~Fendt$^{2}$ \thanks{E-mail: fendt@mpia.de (CF)},
Dominik R.G.~Schleicher$^{1}$,
\newauthor
Javier~Lagunas$^{1}$,
Javier~Pedreros$^{1}$,
Neil~ M. Nagar$^{1}$,
Felipe Agurto$^{1}$,
\\
$^{1}$Departamento de Astronom\'ia, Facultad Ciencias F\'isicas y Matem\'aticas, Universidad de Concepci\'on, \\
Av. Esteban Iturra s/n Barrio Universitario, Casilla 160-C, Concepci\'on, Chile\\
$^{2}$Max Planck Institute for Astronomy, K\"onigstuhl 17, D-69117 Heidelberg, Germany\\
}
\date{Accepted XXX. Received YYY; in original form ZZZ}

\pubyear{2022}

\begin{document}
\label{firstpage}
\pagerange{\pageref{firstpage}--\pageref{lastpage}}
\maketitle


\begin{abstract}

The Event Horizon Telescope Collaboration (EHTC) has presented first - dynamic-range limited - images of the black hole shadows in M87 and Sgr A*. The next generation Event Horizon Telescope (ngEHT) will provide higher  sensitivity and higher dynamic range images (and movies) of these two sources plus image at least a dozen others at $\leq$100 gravitational radii resolution. We here perform an exploratory study of the appearance of winds and jets in such future observations. To do this we use  M87 and Sgr A* as reference systems: we do not aim to exactly reproduce them, but rather to determine how their observed images will depend on specific physical assumptions. Even in the case of similar or the same dynamics, the images depend significantly on global parameters such as the black hole mass and the mass accretion rate. Our results provide guidance in the interpretation of future high-resolution images, particularly if a wind or jet is detected.
\end{abstract}



\section{Introduction}

Astrophysical black holes (BHs) are found with masses ranging from a few solar masses to supermassive BHs of masses greater than $10^{6} M_{\sun}$. BHs initially were a basic outcome of Einstein's General Theory of Relativity (GR) defined by a one-way causal boundary in space time also known as the 
event horizon from which even light cannot escape, making the direct detection of BHs impossible. 
BHs can be detected only indirectly by their gravitational influence on the surrounding matter and/or space-time, by their accretion which results in often luminous emission from radio to $\gamma$-rays,  or by gravitational wave emission.
The Event Horizon Telescope (EHT) has presented first images of the gravitionally-lensed ring, and BH shadow, of the supermassive BH of M87 \citep{EHTC2019a,EHTC2019b,EHTC2019c,EHTC2019d,EHTC2019e} in the heart of Virgo cluster, and of SgrA* at the center of the Galaxy \citep{,EHTC2019f,EHTC2022a,EHTC2022b,EHTC2022c,EHTC2022d,EHTC2022e,EHTC2022f}.

Active galactic nuclei (AGN), accretion powered supermassive BHs in the nuclei of some galaxies, have intrigued astronomers for ages as being the most powerful sources in the sky. 
They can outshine even the entire stellar population of the galaxy which hosts them such as in quasars which are among the most luminous sources in the sky \citep{Schmidt1963, Sanders1989, Novikov1973}.
These are believed to be powered by high accretion rates on to the supermassive BH through a geometrically thin but optically thick disk \citep{Shakura1973, Sun1989}. 
On the other hand, AGN in the local universe, including M87 and SgrA*, dominantly show lower accretion rates onto the central BH  \citep{Ichimaru1977, Narayan1995, Blandford1999, Nagar2005, Yuan2014}. 

Many AGN display signatures of outflows and large-scale jets which are observed across the spectrum. 
Jets appear as collimated structures launched at relativisitic velocities from very close to the BH. 
The EHT has recently imaged the innermost jets of 3C 279, CenA, and the blazar J1924-2914 \citep{EHTC2020,Janssen2021,Issaoun2022} with 20 $\mu \rm as$ resolution paving a way to observe and understand the physical processes involved in launching jets around supermassive BHs. 
Jets may either be powered by magnetic fields threading the event horizon, extracting the rotational energy of the BH, \citep{Blandford1977} or by the magneto-rotational acceleration of matter from the accretion flow \citep{Blandford1982}. 
The presence of strong magnetic fields close to the event horizon thus leads to synchrotron radiation which peaks between the radio and the far-infrared \citep{Ho1999} and could originate from the accretion flow, jet base or disk wind.

The first ever images of the BH shadows of M87 and SgrA* have paved the way for understanding the physical processes around supermassive BHs. 
While both sources have roughly the same flux and angular size in their gravitationally lensed ring, they are individually interesting as their masses, and potentially spins, are on the two extreme ends of the mass and spin spectrum of supermassive BHs. 
Although the BH mass of M87 ($M_{BH}=6.5 \times 10^{9}~M_{\odot}$) is three orders of magnitude larger than that of SgrA* ($M_{BH}=4.16 \times 10^{6}~M_{\odot}$), it is also at a distance approximately 3 orders of magnitude farther ($D_{M87}=16.8$ Mpc) than SgrA* ($D_{SgrA*}=8.127$ kpc) [see for e.g. \citet{EHTC2019a, EHTC2022a}]. This gives rise to interesting observational similarities such as their lensed photon rings project similar angular sizes on earth ($39~\mu$arcsec for M87 and $52~\mu$arcsec for SgrA*), which make them the most interesting targets for the EHT. Also, they both display similarities in the flux received through our telescopes. The total accretion power of M87 is estimated to be typically $\sim~10-100$ times larger than SgrA* \citep{EHTC2021}. The higher power of M87 could mean that it is being fed directly from a larger reservoir of gas than the stellar winds which are expected to feed SgrA*. This may be the cause for another significant difference between the sources i.e. the prominent, powerful jet launched from M87, which can be observed at multiple wavelengths and across nearly eight orders of magnitude in size \citep{EHT2021MWLG}. The extended one-sided M87 jet, and the relative faintness of the nuclear counter-jet, provides strong constraints on the orientation of the jet, and thus the spin axis of the black hole: an inclination of $\sim 20\degr$ to the line of sight was used for numerical simulations of M87 \citep{EHTC2019e}.  For SgrA* there are no strong constraints on the spin axis yet, although the model fits from the EHTC suggest an inclination angle of $30\degr$ to the line of sight \citep{EHTC2022a}). These interesting characteristics of M87 and SgrA* make them the best candidates for testing different theoretical models.  

The Event Horizon Telescope, in science operation since 2017, combines a network of 8 to 10 globally distributed mm-wave telescopes using the technique of very large baseline interferometry (VLBI). The next generation EHT (ngEHT), currently in  the design stage, aims to double this number of telescopes, double bandwidth (thus sensitivity), and implement dual 230/345 GHz observing. This will allow significant improvement in image quality, thus observation of finer structures and time-lapse movies of horizon scale structure in M87 and SgrA*, and the imaging of additional sources at better than 100 gravitational radius resolution. On the longer term the spatial resolution can be bettered either using higher frequency (e.g., 700 GHz) on the ground, or longer baselines via ground-satellite interferometry. Resolutions can be significantly improved with earth-Geostationary (GEO-VLBI) or earth-L2 (L2-VLBI) baselines. In addition to improving the image quality with extended baselines, ngEHT also plans to improve on image construction algorithms from the data to be able to obtain the the strongest possible constraints on the complex structures around the BH event horizon.

Given the limits of the 2017 EHT intrinsic resolution and dynamic range, it has been difficult to address some questions such as the whether the emission signatures are from the accretion disk or disk wind or jet base, whether the ring features are in complete agreement with standard GR or if there is a possibility to constrain the spin and accretion rate accurately. The ngEHT  will already be an improvent with better image fidelity (more $u-v$ coverage) and sensitivity with a resolution gain from moving to 345 GHz as well as super-resolution techniques. However,we expect that the longer base-lines of future earth-space based VLBI  (a possibility through space based ngEHT) will result in an enhancement in the resolution which will allow us to distinguish between various physical scenarios. Through this work we intend to demonstrate how to distinguish certain theoretical models with current and future VLBI probes, given the mass and flux constraints for two systems with contrasting mass and spin. While here our aim is not to reproduce specific images, we will take some guidance from the well-studied systems M87 and Sgr A*, with the aim of motivating the properties of a typical source. For this purpose, we will consider the black hole masses and the reported spectral energy distribution (SED) information at low frequencies of these two sources reported in the literature [for reference see\citep{Narayan1998, Prieto2016}] to derive some reference systems that have similar emission properties. Based on these reference systems (M87 or SgrA*), we will study the possibility to identify signatures of winds and jets in future observations. The paper is organised as follows: In section [\ref{subsec:dyn}] we describe briefly the dynamical set up of our simulations while in section [\ref{subsec:raytr}] the post-processing and ray-tracing methods are discussed especially in the context of the ones chosen for this work. In section [\ref{sec:spec}], we discuss how the accretion rates are determined from the synthetic SED which fit to the observed data for our simulations for M87 and SgrA* like systems while in section [\ref{sec:Insten}], we discuss the features obtained in the intensity maps and profiles with the parameters obtained from the spectral fitting. A comparision of the deduced results for a M87 and SgrA* like system is discussed in section [\ref{sec:Comp}]. Finally in section [\ref{sec:Conclu}] we provide the primary insights obtained from this investigation.

\section{Methodology}


\begin{table*} 
\begin{center}
\caption{Characteristic parameters of our simulation runs with the Eddington ratios obtained from SED fits to the data for M87 as tabulated in \citet{Prieto2016} and and for SgrA* in \citet{Narayan1998}. Shown are the simulation run ID, followed by the simulation inputs which are, the initial disc plasma beta $\beta_0$ (at the inner disc radius),
the maximum floor density $\rho_{\rm flr}$ and internal energy $u_{\rm flr}$ (both following a power law in $r$), the Kerr parameter of the BH $a$, the radius of the ISCO $R_{\rm isco}$, and specific comments on the simulation runs, respectively. Following the simulation inputs, the two columns show the Eddington ratios obtained from these simulations for a M87-like and SgrA* like system with SED fits. A background diffusivity of $\eta_{\rm back}=10^{-3}$ is applied, as well as a maximum disc magnetic diffusivity $\eta_{\rm max}=0.01$. We generally apply a density contrast $\rho_{\rm cor} / \rho_{\rm disc} = 0.001$ at the inner disc radius.}
\label{tab:para_dynamics}
\begin{tabular}{ccccccccl}
\hline
\hline
\noalign{\smallskip}
\multirow{2}{*}{$ID$} & \multicolumn{5}{c}{Simulation Inputs} & \multicolumn{2}{c}{$\dot{m}$ from SED fits} & \multirow{2}{*}{$Comments$} \\
 & $\beta_0 $ & $\rho_{\rm flr}$ & $u_{\rm flr}$  & $a$ & $R_{\rm isco}$ & $\dot{m}_{\rm th-M87}$ &  $\dot{m}_{\rm th-SgrA*}$ & \\
\noalign{\smallskip}
\hline
\noalign{\smallskip}
\noalign{\smallskip}
 20EF  & 10  & $10^{-5}$ & $10^{-8}$  & 0.9375 & $2~R_{\rm g}$ & $2.0 \times 10^{-5}$ & $5.6 \times 10^{-7}$ & reference run, similar to \citet{Vourellis2019} \\
 21EF  & 10  & $10^{-4}$ & $10^{-6}$  & 0.9375 & $2~R_{\rm g}$ & $1.0 \times 10^{-5}$ & $5.0 \times 10^{-7}$ & as 20EF, floor density higher \\
 22EF  & 10  & $10^{-4}$ & $10^{-6}$  & 0.0    & $6~R_{\rm g}$ & $7.0 \times 10^{-5}$ & $5.0 \times 10^{-6}$ & as 20EF, $a=0$, infall to BH, disc wind \\
 23EF  & 10  & $10^{-3}$ & $10^{-5}$  & 0.9375 & $2~R_{\rm g}$ & $3.0 \times 10^{-6}$ & $1.5 \times 10^{-7}$ & as 21EF, floor even higher, no BZ \\
 24EF  &  1  & $10^{-3}$ & $10^{-5}$  & 0.9375 & $2~R_{\rm g}$ & $4.0 \times 10^{-7}$ & $2.0 \times 10^{-8}$ & as 21EF, $\beta_0$ lower, stronger magn. field \\
 26EF  & 0.1 & $10^{-3}$ & $10^{-5}$  & 0.9375 & $2~R_{\rm g}$ & $2.0 \times 10^{-7}$ & $1.0 \times 10^{-8}$ & as 24EF, $\beta_0$ even lower, even stronger magn. field \\
  \noalign{\smallskip}
 \hline
 \noalign{\smallskip}
 \end{tabular}
 \end{center}
\end{table*}

Our approach follows a two-step procedure. First, we carry out resistive general relativistic magneto hydro dynamic (GR-MHD) simulations of the disk-jet evolution for a range of parameters. Secondly, we post-process these dynamical data applying ray-tracing procedures. Note that the dynamical data are scale-free by definition as these are pure magneto hydro dynamic (MHD) simulations. We thus need to re-scale our physical parameters, such as mass density, gas pressure, and magnetic field strength to astrophysical units.
In fact, this approach allows us to apply our dynamical models to accreting BH systems with various physical conditions such as BHs of different mass, accretion rates and magnetic flux. In turn, we may derive these parameters of the BHs by comparing our results to observed systems.

\subsection{Dynamical Modeling}\label{subsec:dyn}
The GR-MHD models we use here are the same as in \citet{Bandyopadhyay2021}.
Here, we summarize their main features and the numerical setup of the simulations for convenience. 

We have applied the resistive GR-MHD code rHARM3D
\citep{Qian2017, Qian2018, Vourellis2019} 
that is based on the well-known HARM code 
\citep{Gammie2003, McKinney2004, Noble2006,Noble2009}.
The application of a physical resistivity is essential for such simulations as it allows both for a long-term mass 
loading of the disk wind and jet, and for a smooth accretion process.
For such simulations, it is believed that physical resistivity is essential as it allows both for a long-term mass loading of the disc wind and jet, and for a smooth accretion process.
This is a consequence of the involved magnetic diffusivity.
In addition, this allows to treat reconnection in a way that is physically well defined.

The GR-MHD simulations considered here are axisymmetric on a spherical grid, implying that the vector components of all 3 dimensions are considered,
however, any derivatives in the $\phi$ direction are neglected partly due to the exceptionally high CPU demand in case of resistive MHD.
However, we believe that our approach is fully sufficient in order to investigate the radiative signatures of the
black hole-disc-wind-jet system. Of course, any 3D features like instabilities in the disc or in the jets cannot be treated,
nor any orbiting substructures can be found in these components.
For our qualitative investigation, these assumptions are sufficient.

It is worth summarizing the differences of our dynamical model setup in comparison to other attempts to model intensity maps and 
spectra from GR-MHD simulations such as e.g. published by \citet{Dexter2012, Moscibrodzka2016, EHTC2019e}.
We work with resistive GR-MHD, which allows for smooth disc accretion and launching of a disc wind. Our disc is thin, in Keplerian rotation, and is threaded by a large-scale magnetic flux. This allows to drive strong disc winds (that, on larger spatial scales, are supposed to evolve into a jet). In these cases, it is the magnetized disc wind that is more efficient in angular momentum removal due to its large lever arm.

A common feature of all GR-MHD simulations is the application of a so-called floor density set as a lower density
threshold in order to keep the MHD simulation going.
This numerical necessity strongly affects the resulting dynamics of the axial Blandford-Znajek (BZ)
jet --  the mass load carried away with the jet and the jet velocity \citep{Qian2018, Vourellis2019}. 
On the other hand, the disk wind dynamics is governed by the mass loading from the disk,
which is self-consistently derived from MHD principles.

Our simulations follow the setup of \citet{Vourellis2019}.
The output of our GR-MHD simulations comes in normalized units where $G=c=M=1$.
In particular, all length scales used in these simulations are in gravitational radii $R_{\rm g}$, while velocities 
are normalized to the speed of light.
All physical variables need to be properly re-scaled for post-processing of radiation.

A further discussion of our model approach including the parameter space involved, as well as maps of the essential dynamical variables, can be found in \citep{Bandyopadhyay2021}.
In brief,
we investigate (i) a reference simulation labeled as {\em 20EF} which applies similar simulation parameters as \citet{Vourellis2019},
(ii) a  simulation {\em 21EF} with a (ten times) higher floor density, thus a higher mass loading of the BZ jet, 
(iii) a simulation {\em 22EF} with a Kerr parameter $a=0$,
(iv) a simulation {\em 23EF} with an even higher floor density (another factor ten) resulting in the absence of BZ jet, 
(v) a simulation {\em 24EF} with a lower plasma-$\beta$ (factor ten), resulting in a stronger magnetic flux, and 
(vi) a simulation {\em 26EF} with an even stronger magnetic flux (another factor ten).

The differences in the model setups are mainly in terms of the magnetic field strength, BH rotation and floor density.
However, these parameter have an immediate impact on the jet dynamics, namely the jet speed (Doppler boosting), and mass flux. 
In combination with the magnetic field, the synchrotron emission is also affected.
Note, however, that the latter also depends on the scaling of the simulation variables in astrophysical units.

The accretion rate given in the GR-MHD simulations is in code units, $\overline{\dot{M}} \simeq 0.1$,
which can be re-scaled to astrophysical units. As an example typical AGN accretion rates of 
\begin{equation}
    \dot{M} \equiv \rho_0  R_{\text g}^2  {\text c} \overline{\dot{M}} \simeq 10^{-3}\,M_{\odot}\text{yr}^{-1} \label{eqn:acc}
\end{equation}
constrain the (maximum) disk density to
\begin{equation}
\rho_0 = 4\times 10^{-15}\frac{\text{g}}{\text{cm}^3}  \left[ \frac{ \dot{M} }{10^{-3}\,M_{\odot}/\text{yr}} \right] 
                                    \left[ \frac{ M_{\text{BH}} }{ 5\times 10^9 M_{\odot} } \right]^{-2}
                                    \left[ \frac{ \overline{\dot{M} } }{ 0.1 } \right]^{-1}\!\!\!\!
\end{equation}        
\citep{Vourellis2019}.
The re-scaled astrophysical density is $\rho = \rho_0 \bar{\rho}$, where $\bar{\rho}$ in code units follows from 
our simulations.
For illustration, for the example of M87 one can use the values provided by the EHT collaboration \citep{EHTC2019b}
with an accretion rate of 
$\dot{M}\sim 2.7\times 10^{-3}\,M_{\sun}\rm yr^{-1}$ 
and a BH mass of $M = 6.2\times10^9\,M_{\sun}$.

\begin{figure*}
 \begin{center}
 \includegraphics[height=3in,width=3in]{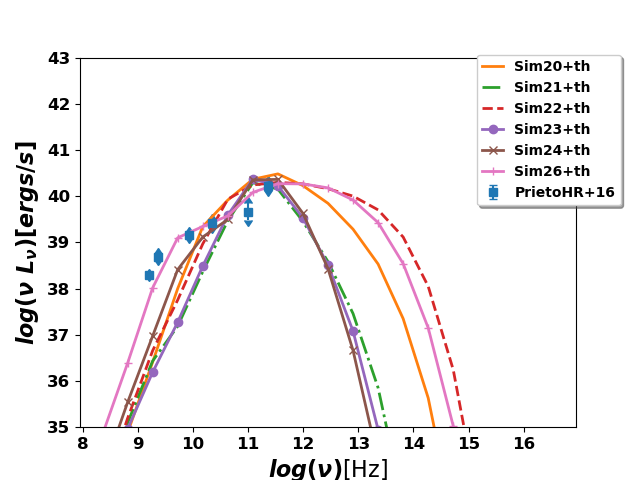}
 \includegraphics[height=3in,width=3in]{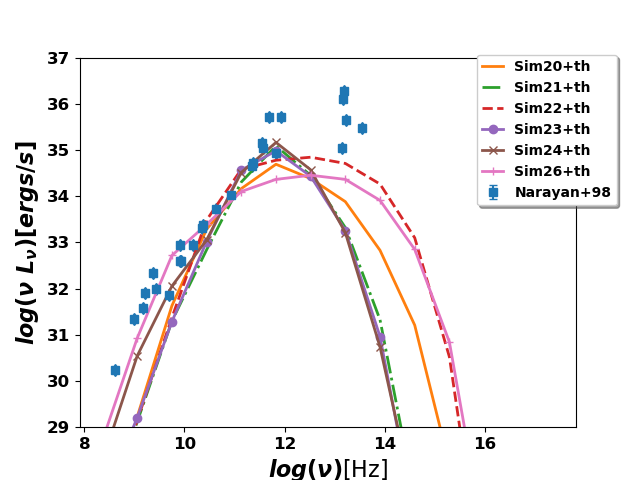}
  \caption{The synchrotron spectra dominated by thermal electrons for M87 (left) and SgrA* (right) like systems. The Eddington ratios obtained for each of the simulations fitted to the data for M87-like system \citep{Prieto2016} and SgrA*-like system \citep{Narayan1998} are shown in the columns 7 and 8 of Table [\ref{tab:para_dynamics}]}
  \label{fig:spectrathM87SgrA}
\end{center}
\end{figure*}

\subsection{Parameter fitting and ray-tracing}\label{subsec:raytr}
GR-MHD codes like the one we have used in this work 
typically apply normalized units where $G = c = M = 1$. These simulations need to be post processed with physical parameters such as black hole mass, accretion rates etc., to obtain a physical realization of an AGN. Here we post process our simulations for the BH masses of M87 and SgrA* and vary the accretion rate (fixing other parameters) to obtain synthetic SED comparable to the observed data in the radio bands for these sources. The post processing of these GR-MHD simulations are done using ray tracing codes.
 
The image of a BH is characterized by a "shadow" region towards which no photons arrive due to the curvature of space-time around the BH; the unlensed "photon ring" ($3~R_g$), corresponding to bound photon orbits around the BH; and as a result of lensing, the photons from the inner part of the disk appear somewhat further outside that region leading to lensed ring marked by inner ($5.2~R_g$) and outer ($6.2~R_g$) boundaries \citep{Gralla2020a, Gralla2020b, Gralla2020c, Gralla2019, Johannsen2013, Beckwith2005, Luminet1979, Bardeen1973}. These features are a result of strong GR effects around the vicinity of the BH. The use of a ray-tracing code is to post process the simulated dynamical model to calculate the observed intensity on locations (pixels) of an observer’s camera for a given model of emission and absorption. Here we use the publicly available general relativistic radiative transport code GRTRANS \citep{Dexter2016}. Within GRTRANS, the Boyer–Lindquist coordinates of the photon trajectories are calculated from the observer towards the BH (i.e. the rays are traced) for each pixel in the camera. The observed polarization basis is then parallel-transported into the fluid frame. The local emission and absorption properties at each point are calculated. The radiative transfer equations are solved along those rays. We refer to the original paper by \citet{Dexter2016} for details on the working of the code.

The jet and wind dominated regions are highly magnetized and thus it can be assumed that synchrotron emission is the most dominant emission mechanism in these regions \citep{Yuan2014}. In this work we consider synchrotron emission from a population of thermal electrons as the primary emission process.
This emission, often referred to as thermal synchrotron, depends on the thermal state of the electrons where the electron temperature should ideally be determined from the plasma physics. However, currently setting up simulations of a fully functional plasma dynamics in a GR set up is computationally expensive. In the absence of such a physically complete model, an alternate prescription is applied which replicates the plasma temperature model. In this scenario, the
gas temperature at each point in the grid is determined from the
internal energy at those grid points assuming an ideal gas prescription. The electron temperature is deduced from the gas temperature assuming that the plasma consists only of ionised hydrogen. As prescribed by \citet{Moscibrodzka2009,Moscibrodzka2016}, the electron temperature is approximated from the following relation as

\begin{eqnarray}
    T_{e}&=&T_{\rm gas}/(1+R), \nonumber\\
    R & = & \frac{T_{\rm p}}{T_{\rm e}} = R_{\rm high}\frac{b^2}{1+b^2}+R_{\rm low}\frac{1}{1+b^2},
\label{eqn:temp}
\end{eqnarray}
with $b=\beta/\beta_{\rm crit}$, $\beta=P_{\rm gas}/P_{\rm mag}$ and $P_{\rm mag}=B^2/2$. The value of $\beta_{\rm crit}$ is assumed to be unity, and $R_{\rm high}$ and $R_{\rm low}$ are the temperature ratios that describe the electron-to-proton coupling in the weakly magnetized ({\it e.g. }discs) and strongly magnetized regions ({\it e.g.} jets), respectively. The emissivity for a given distribution function of electrons is given as
\begin{equation}
    j=\int_{1}^{\infty}d\gamma N(\gamma)\eta \label{eqn:RT},
\end{equation}
where 
\begin{equation}
    \eta=\frac{\sqrt{3}e^2}{8\pi c}\nu_{B}sin \theta_{B}H(\nu,\theta_B)
\end{equation}
is the vacuum emissivity \citep{Melrose1980} with $e-$ the electron charge, $c-$ the speed of light, $\nu_B=\frac{eB}{2\pi mc}$, $H=F\left(\frac{\nu}{\nu_c}\right)$, $\nu-$ the emitted frequency, $\gamma-$ the electron Lorentz factor, $\nu_c=(3/2)\nu_B \sin(\theta_{B})\gamma^2$ and $F(x)=x\int_{x}^{\infty}dyK_{5/3}(y)$ is an integral of the modified Bessel function. For a thermal distribution the distribution function in eqn.[\ref{eqn:RT}] is
\begin{equation}
    N(\gamma)=\frac{n\gamma^2\beta \exp(-\gamma/\theta_e)}{\theta_e K_2(1/\theta_e)} \label{eqn:DFTher},
\end{equation}
where $n$ is the number density of electrons and $\theta_e=\frac{kT_{\rm e}}{mc^2}$ is the dimensionless electron temperature. 

As shown by \citet{Bandyopadhyay2021}, the post-processing parameters which significantly affect the synthetic SED and emission maps are the BH mass and the accretion rate. The accretion rate and black hole mass scale the density, pressure and the true magnetic field from the simulations. The temperature of the gas is obtained from the density and pressure assuming an ideal gas equation as already mentioned. The temperature and the magnetic field are the important physical parameters which affect the synchrotrom emission through the distribution function (eqns. [\ref{eqn:RT}] and [\ref{eqn:DFTher}]).

In this work, we have used the high resolution spectral data from \citet{Prieto2016} for M87 and that from \citet{Narayan1998} for SgrA*. 
Modeled spectra from the simulations are generated by fixing all the other parameters (except for the black hole mass and accretion rate) to constant values for the two systems for simplicity. As this is a qualitative study about the emission features around supermassive BHs, the small changes in the SED due to the inclination angle or the choice of $R_{high}$ and $R_{low}$ in eqn.[\ref{eqn:temp}] are almost insignificant compared to the variation introduced due to the underlying dynamics from the simulations. Once the best fit SED are obtained for each of the simulations, the synthetic images are generated at 230 GHz using the best fit value of the Eddington ratio or the accretion rate for each given BH mass. We use these synthetic images to obtain the intensity profile for several cross-sections of the image.

These synthetic images are further convolved with a Gaussian beam-width to account for the telescope resolution of the EHT and the future ngEHT to obtain secondary 
synthetic images which are expected to be similar to that we observe with the telescopes. 
At 230 GHz, an intrinsic (FWHM) resolution of $\sim20~\mu as$  is achieved for the longest baseline ($D_1\sim 11,000$ Km). Thus using these values, we can obtain approximate values of the resolution for earth space baselines. A Geo-VLBI baseline ($D_2\sim47,000$Km) and a L2-VLBI baseline ($D_2\sim1.5$ million Km) would give us approximate resolutions of $\theta_2\sim 5~ \mu as$ and $\theta_2\sim0.16~ \mu as$ respectively. We assume these resolutions in the sections below (see also  \citet{Pesce2021}).

 
\begin{figure*}
 \begin{center}
 \includegraphics[height=3.5in,width=3.5in]{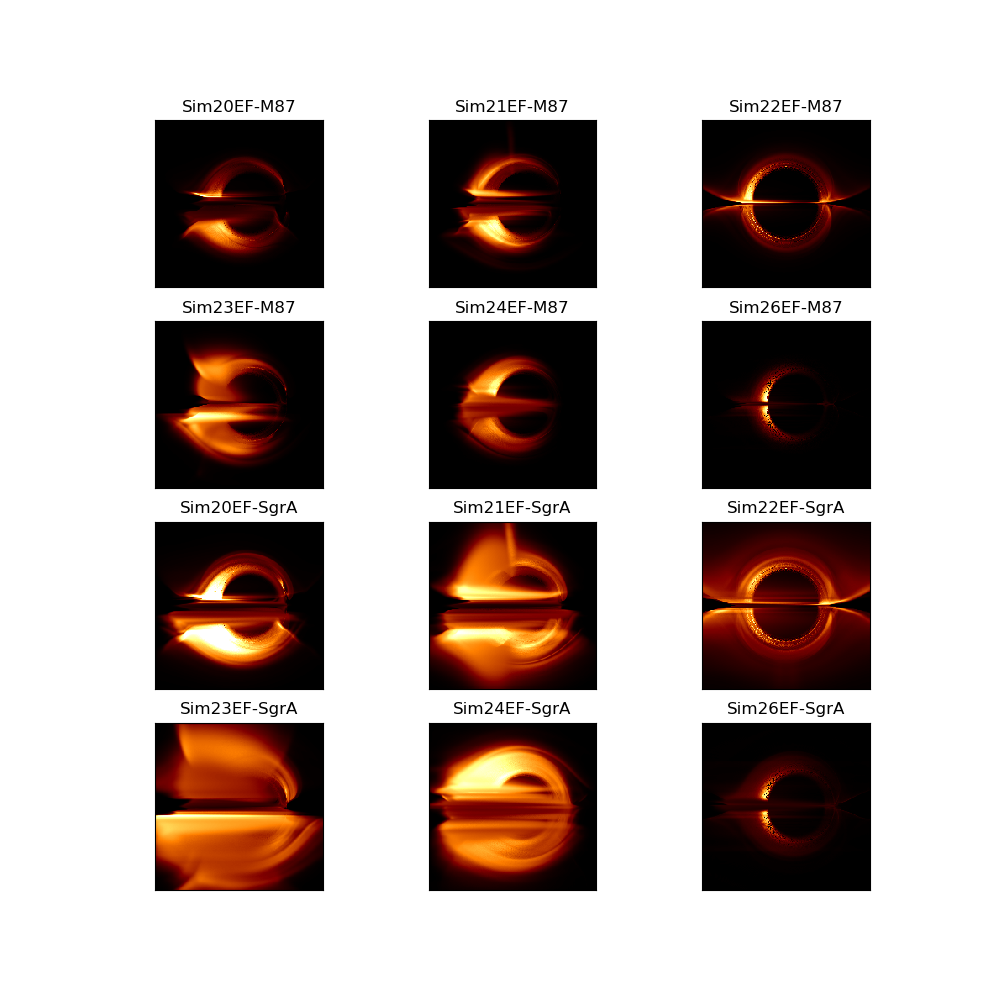}
 \includegraphics[height=3.5in,width=3.5in]{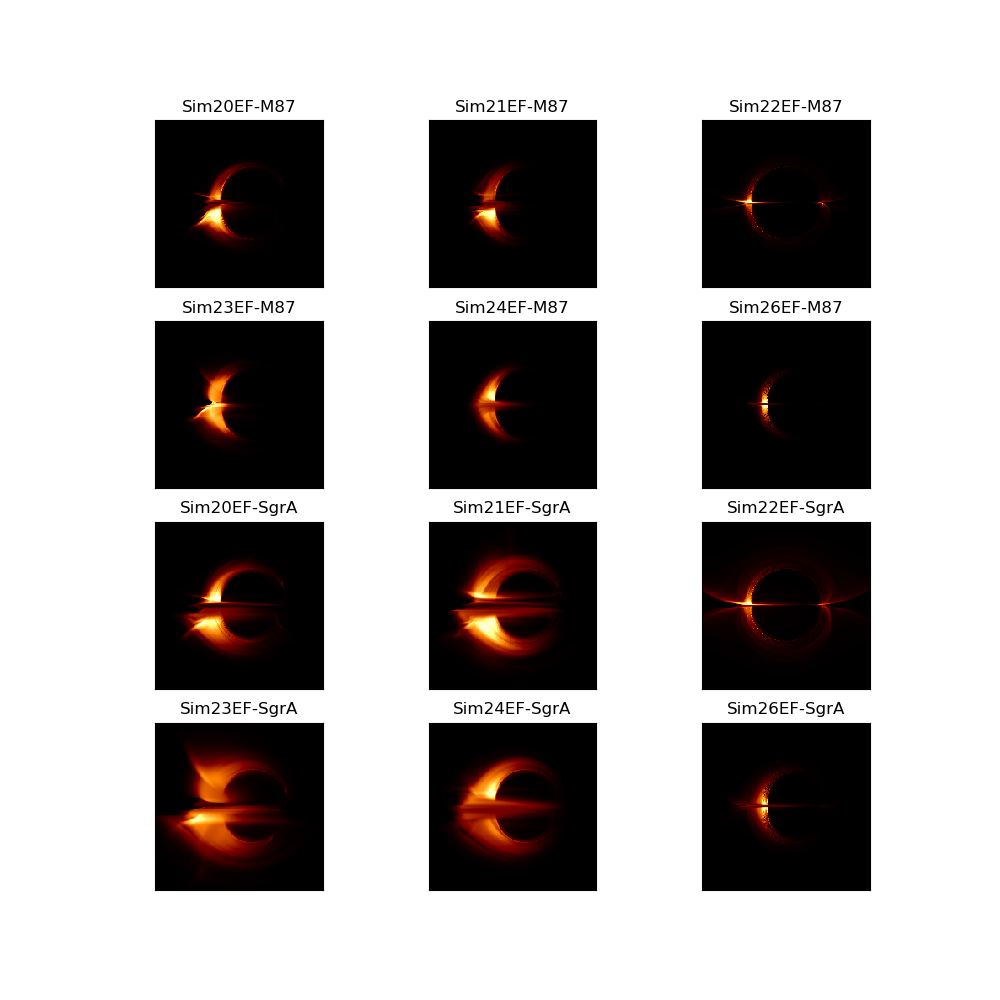}
 \includegraphics[height=3.5in,width=3.5in]{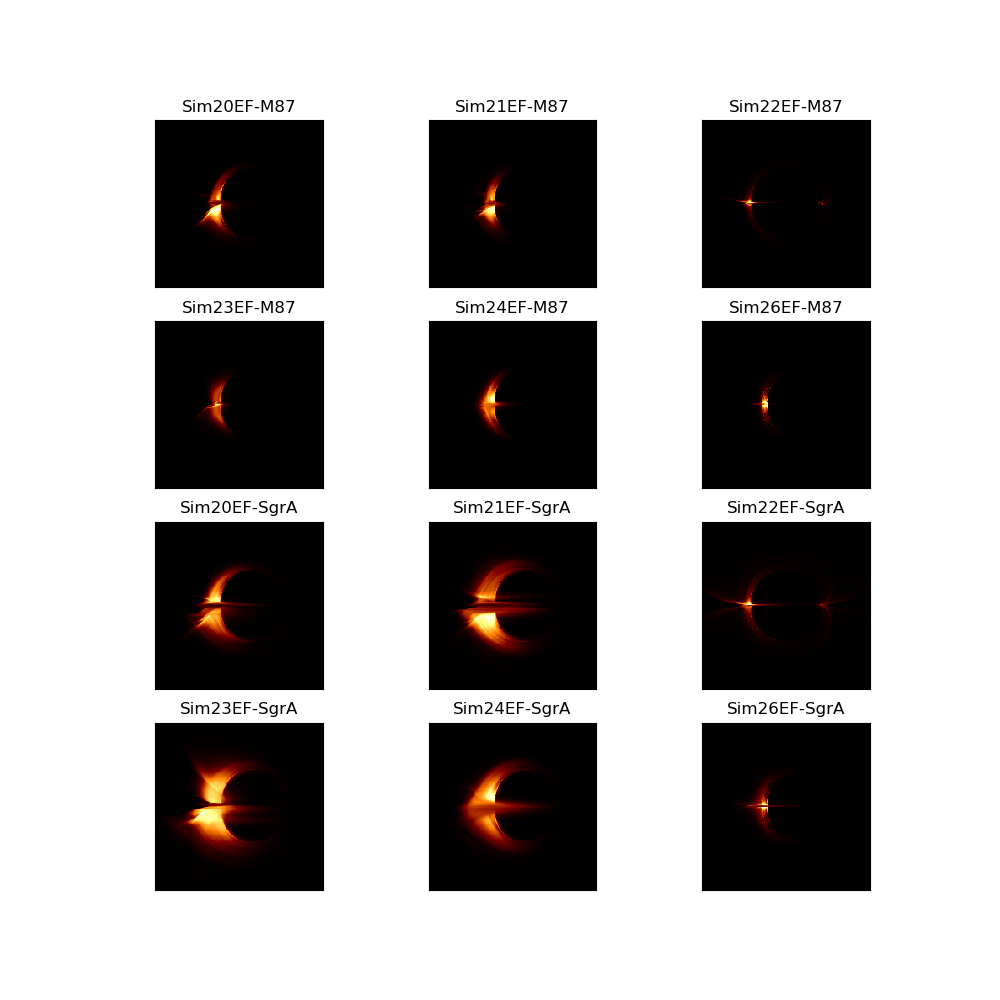}
 \includegraphics[height=3.5in,width=3.5in]{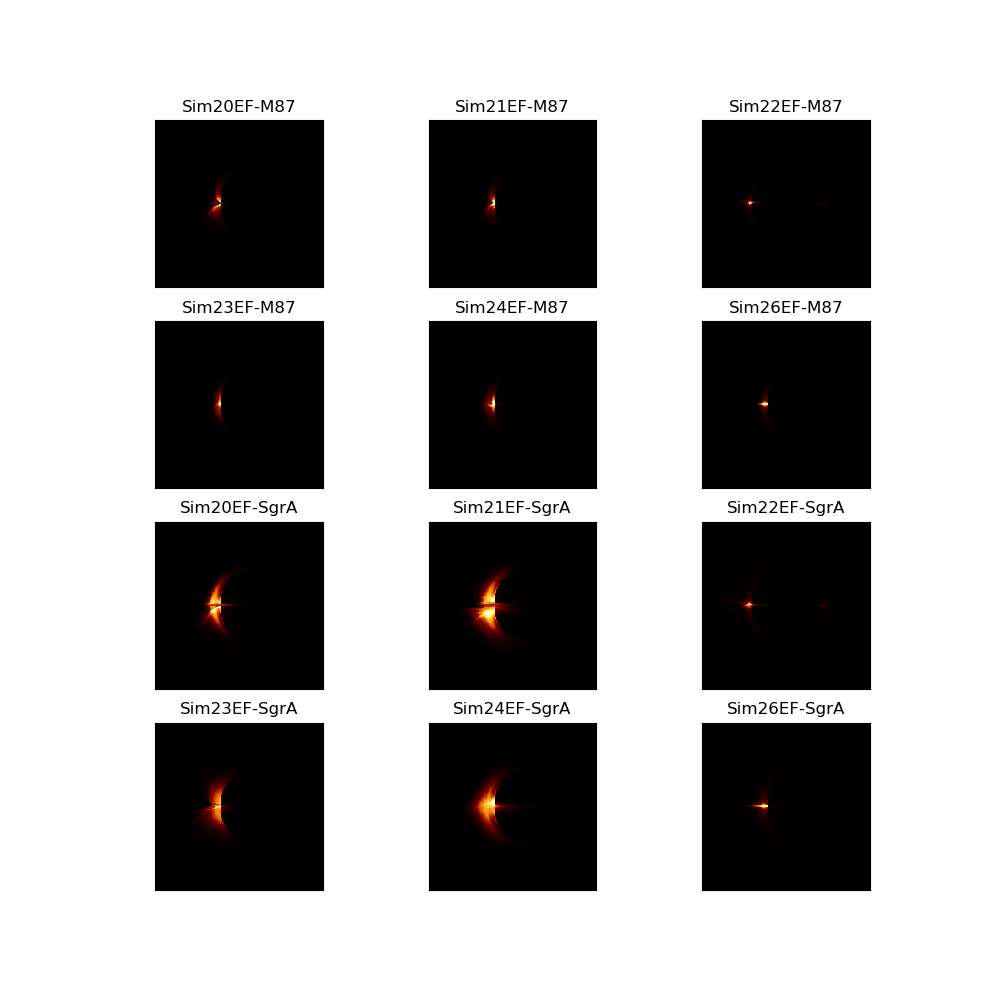}

   \caption{Normalized emission maps for edge-on views for all our simulations for M87 (upper 6 panels in each set) and and SgrA* (lower 6 panels in each set) like systems at 86 GHz (top left), 230 GHz (top right), 345 GHz (bottom left) and 700 GHz (bottom right) obtained with the Eddington ratios mentioned in Table [\ref{tab:para_dynamics}].}\label{fig:M87SgrAedgeon}
\end{center}
\end{figure*}

\begin{figure*}
 \begin{center}
 \includegraphics[height=3.5in,width=3.5in]{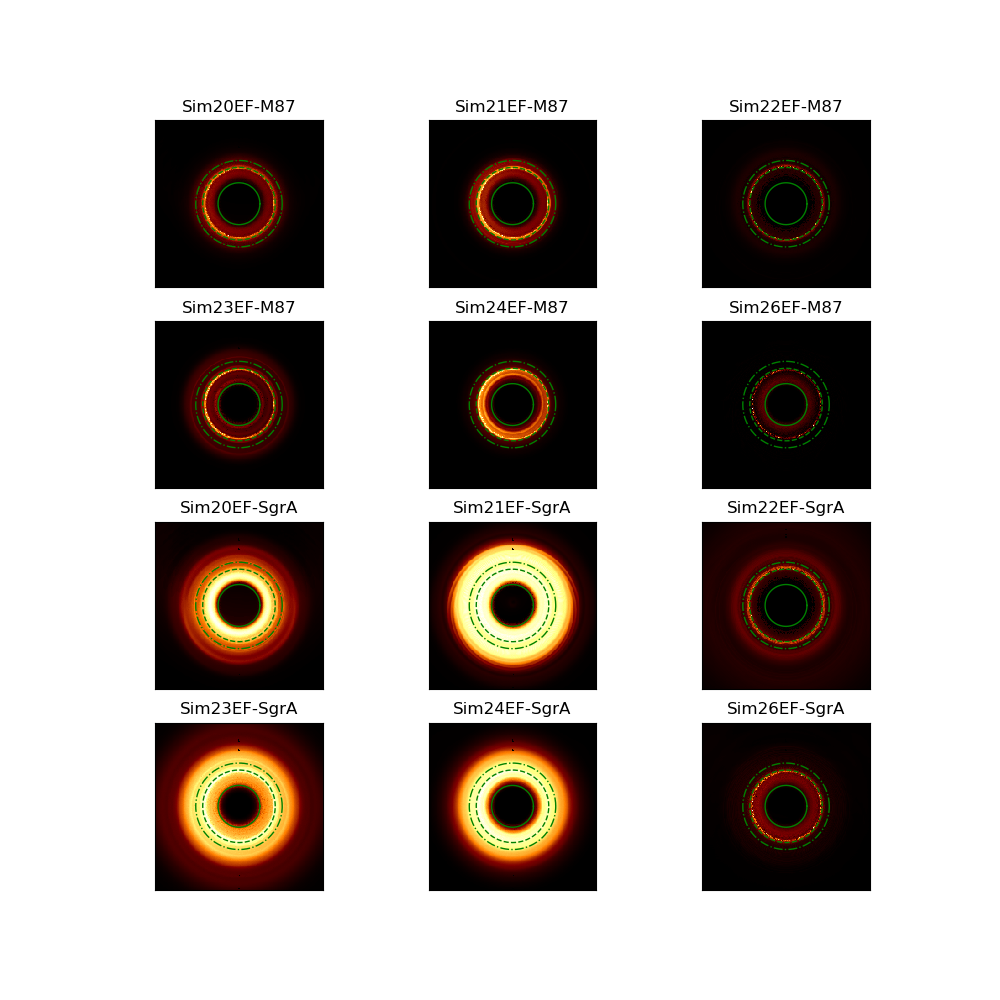}
 \includegraphics[height=3.5in,width=3.5in]{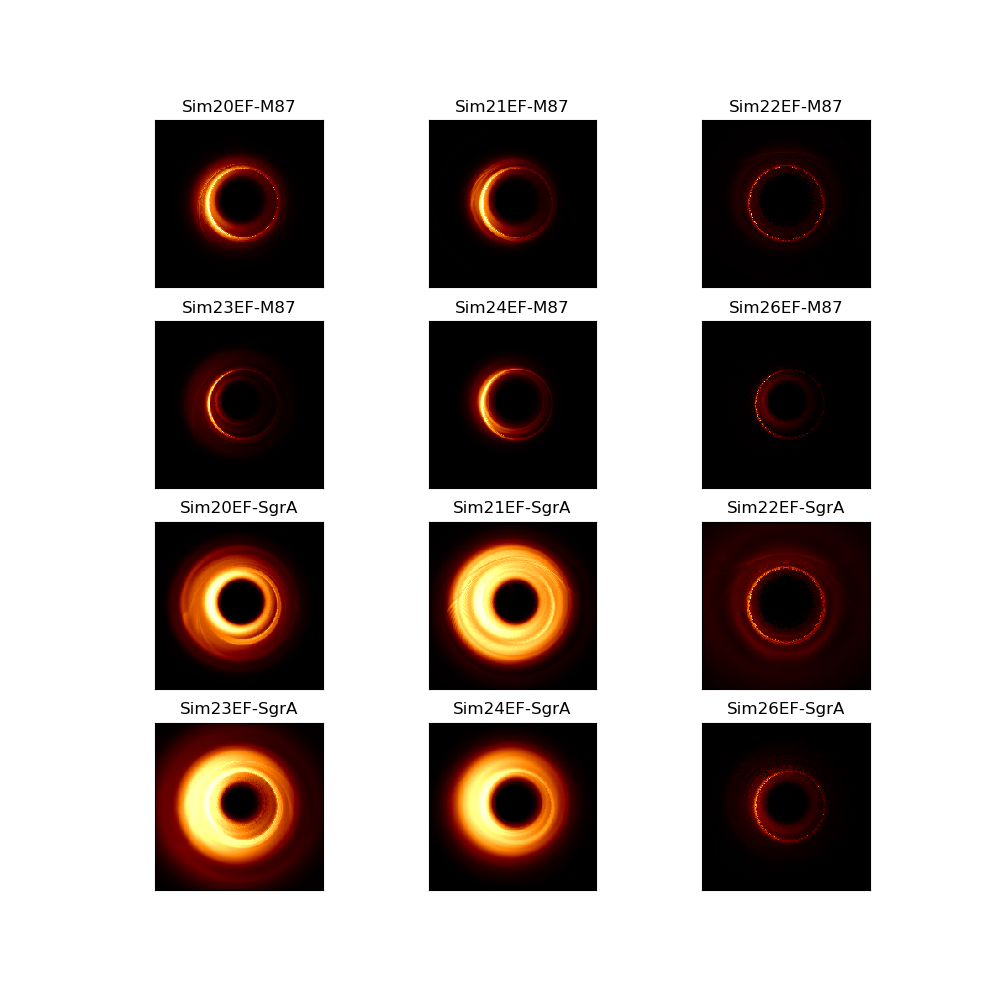}\\ 
 \includegraphics[height=3.5in,width=3.5in]{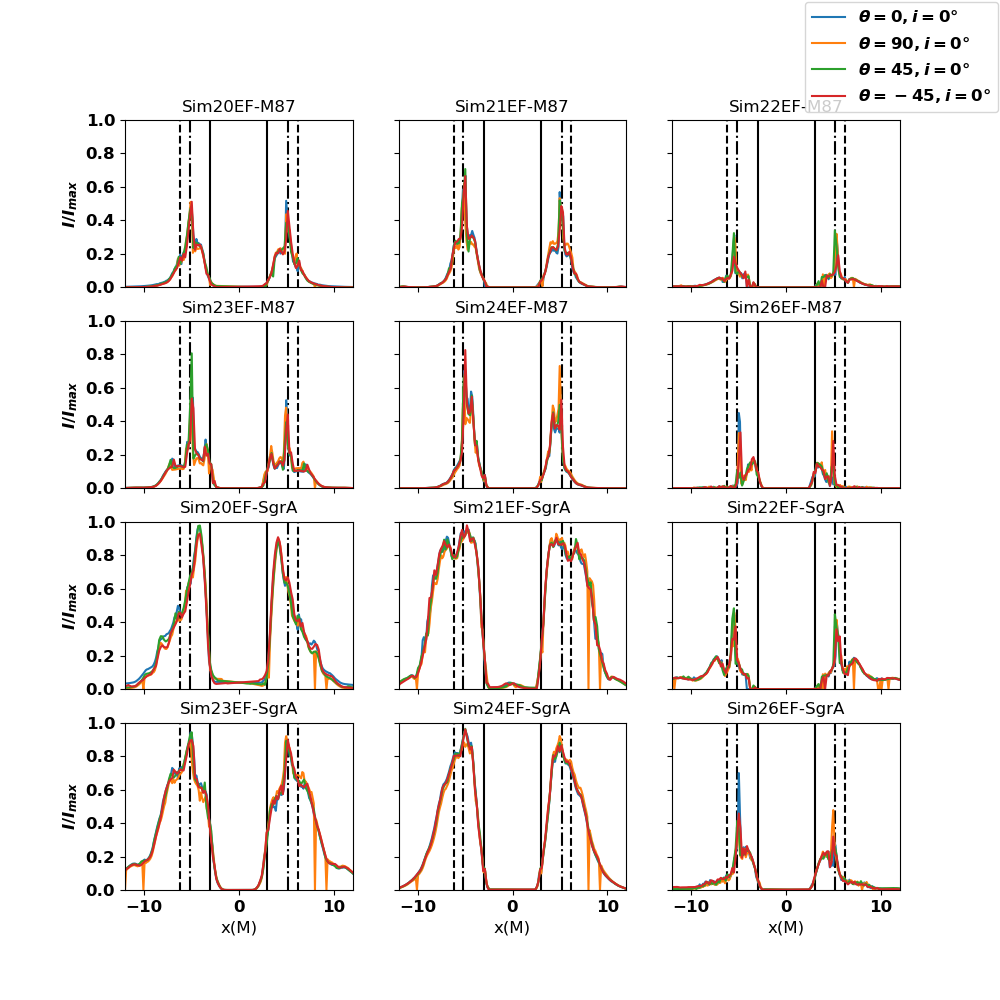} 
 \includegraphics[height=3.5in,width=3.5in]{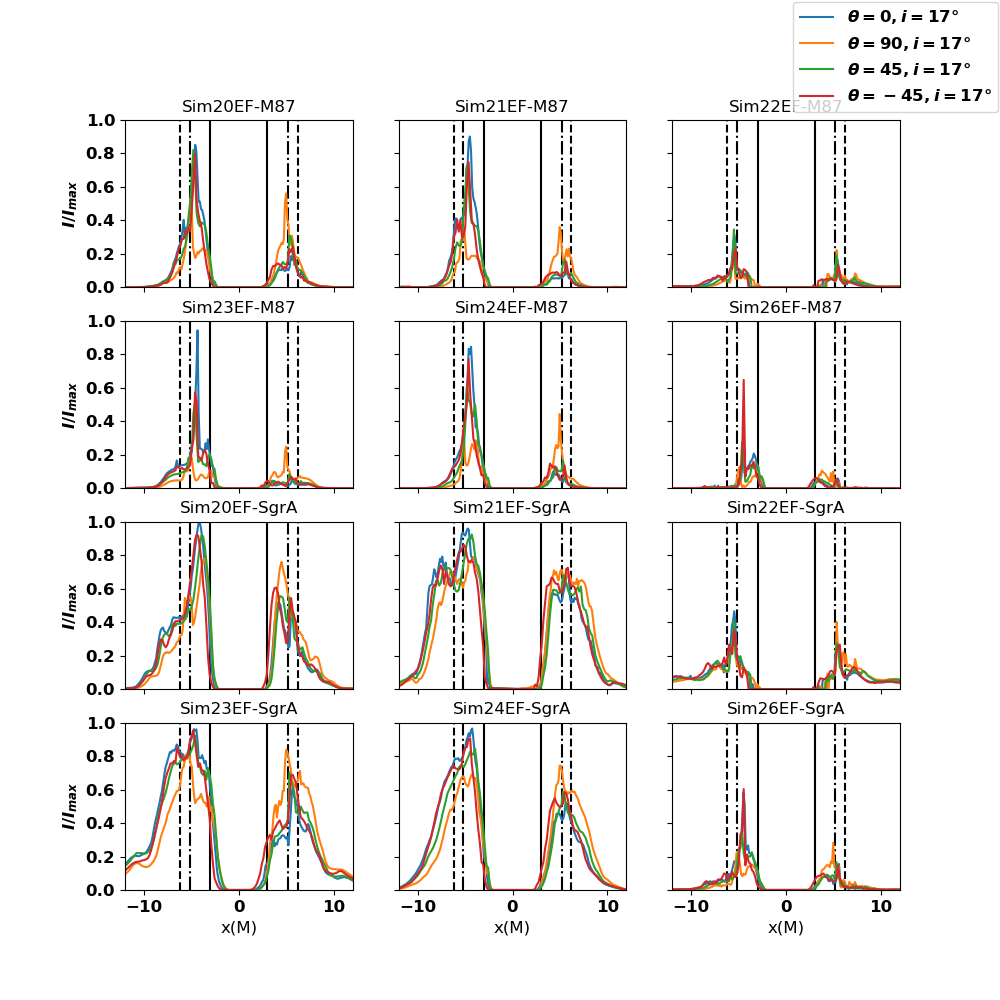}
  \caption{Normalized emission maps (upper) and intensity profiles (lower) (along cross-sections with angles$=0\degr, 90\degr, 45\degr \& -45\degr $ respectively) for all our simulations for M87 (upper 6 in each of the panels) and and SgrA* (lower 6 in each of the panels) like systems with inclination angles $0\degr$or face-on (left) and $17\degr$ (right) for 230 GHz. The emission maps with the face-on view in addition also show the contours of the unlensed photon-ring (3 $R_g$ with continued green circle) and the outer and inner range for the lensed photon ring (5.2 and 6.2 $R_g$ with green dashed circles). The intensity profiles too mark these with black solid line, dot dashed line and dashed line.}\label{fig:M87SgrAang230}
\end{center}
\end{figure*}

\section{Spectral properties}\label{sec:spec}

The primary aim of this work is to investigate the origin of emission signatures from disk winds, jets or the accretion flow for a set of GR-MHD simulations when subject to the known extreme ends of the mass and spin spectrum for supermassive BHs in the local universe. The origin of emission signatures from winds, jets and disk from the different simulations can be distinguished when we can obtain a similar spectral fit from each of them. We thus post processed our simulations with the masses of M87 and SgrA* and varied the accretion rates (by varying the Eddington ratio $\dot{m}=\dot{M}_{\rm acc}/\dot{M}_{\rm Edd}$) for each of our simulations to generate synthetic SED which fit best to the high resolution multi-wavelength data \citep{Prieto2016, Narayan1998} of M87 and SgrA*.  Generally in literature (for reference, see \citet{Moscibrodzka2016, EHTC2019a}, etc.) the SED for the sources of interest are obtained considering the synchrotron emission and its Compton scattering using the GRMONTY code \citep{Dolence2009}. However, in this work, the spectrum obtained with GRTRANS \citep{Dexter2016} using synchrotron emission is sufficient as we are mainly interested in a qualitatively fit of the synthetic SED to the high resolution data in the radio band where we know that synchrotron emission is the most dominant emission process.

M87 has a mass ($6.5 \times 10^9 M_{\odot}$) which is 3 orders of magnitude higher than that of SgrA* ($4.12 \times 10^{6} M_{\odot}$) but is also at a distance (16.9 Mpc) 3 orders of magnitude farther than SgrA* (8.127 kpc) which is in the center of our Galaxy \citep{EHTC2019a, EHTC2022a}. Table [\ref{tab:para_dynamics}] shows that the Eddington ratios (7th and 8th columns) required to match the SED for each of the simulations are higher for M87 compared to SgrA*. This result is similar to previous literature conclusions and implies that M87, the CD galaxy of the Virgo cluster, is fed by a more massive reservoir of gas compared to the trickle from disk winds as speculated for SgrA*. The higher accretion rates in M87 lead to a higher luminosity but since it is also at a distance 3 orders of magnitude farther than SgrA*, both result in similar fluxes.

As can be seen from the spectral fits in Fig.[\ref{fig:spectrathM87SgrA}], {\em 26EF}  (the simulation with spin parameter $~1$ and the lowest value of $\beta$ parameter) fits best to the M87 high resolution data at radio frequencies, while {\em 22EF} (the simulation with spin 0) fits best to the high resolution data for SgrA*. It can be noted here that the accretion rate that fits the M87 data for {\em 26EF} is 2 orders of magnitude smaller than what has been inferred by the EHTC \citep{EHTC2019a}, which could be due to the an intrinsic higher accretion rates ($\dot{M}$ in Eq.~[\ref{eqn:acc}]) in the dynamical simulations used in this work.

It is expected that the SED generated with a power-law, hybrid or Kappa distribution would be the result of accretion rates different from the values deduced in this work. However, this work focuses on the SED fits resulting from a thermal synchrotron emission to deduce the underlying physical parameters from expected observations. 

\section{Intensity maps \& profiles} \label{sec:Insten}
Once the accretion rates are determined by fitting the modeled spectra to the observed data, the intensity maps for the simulations are generated.
The existence of jets and winds if at all present are expected to be detected with a edge-on view as the direction of these flows in general are perpendicular to the accretion flow. However, in reality not all AGN have an orientation with edge-on views. As an example both M87 and SgrA* have very low inclination angles (for M87 this is inferred from  observations of the large scale jet while for SgrA*, this is deduced through model fitting \citep{EHTC2022a}). Hence we first post process our simulations assuming an edge-on view to identify the presence of disk wind and jet signatures and then look for these signatures for lower inclination angles. Fig.[\ref{fig:M87SgrAedgeon}] shows the edge-on views of the post processed images of our simulation for M87-"like" (upper 6 rows of each figure) and SgrA*-"like" (lower 6 rows of each figure) systems at 86 GHz, 230 GHz, 345 GHz and 700GHz respectively. 
Since low frequency photons cannot escape the inner regions due to synchrotron self-absorption, only the outer regions are visibile at lower frequencies and hence the likelihood of the visibility of any signatures of winds and jets at lower frequencies. At higher frequencies such as at 345 GHz, the emission from regions in the vicinity of the black holes event horizon are visible from which the photons with the highest energies are emitted. As can be seen in Fig. [\ref{fig:M87SgrAedgeon}] the emission signatures of jet and disk wind are more prominent at 86 GHz (see for example the emission maps for {\em 21EF} and {\em 23EF} for SgrA* like system) while those same features are not prominent at 345 GHz where the emission is  expected from the inner most regions. However, 230 GHz is an intermediate frequency where both these effects are visible. These images in general are brighter on the left including the ones with jet and disk features  due to Doppler beaming. 

At present it is difficult to obtain good resolution images at 86 GHz
even with an earth-size telescope. As can be observed from Fig.[\ref{fig:M87SgrAedgeon}], 230 GHz is thus the ideal frequency to trace the jet and wind signatures back to horizon scales with the resolutions available in the present. As we move towards higher frequencies, it is difficult to distinguish the traces of any jet or outflow signature distinctly and this is independent of any telescope resolution. Although at higher frequencies, we may able to probe regions in the proximity of the BH event horizon, it may still be difficult to comment on the origin of jets and outflows  at these high energies. This may however change if the accretion rate is increased in which case there will be an overall increase in flux throughout the spectrum (the images for low inclination angle at 345 and 700 GHz are in the Appendix A). Here we, thus, concentrate on investigating emission features of our interest at 230 GHz. The simulations are thus further processed at 230 GHz
for a face-on view and low inclination angle ($17\degr$) as shown in fig.[\ref{fig:M87SgrAang230}]. 

Although from a glance distinguishing emission features from winds and jets is not obvious for low inclination angles or face-on views, however observing the emission features from different simulations especially from their emission profiles hint at the existence of disk winds or jets as shown in fig.[{\ref{fig:M87SgrAang230}}]. As an example one can deduce from the figure that the emission profile for {\em 21EF} is broader than {\em 20EF} for SgrA* like system for both face-on and a small inclination angle of $17\degr$ for a similar Eddington ratio. This comparison although not quite evident from the edge-on view at 230 GHz but the edge-on view at 86 GHz clearly shows a jet signature for  {\em 21EF}. Similarly a large scale wind signature is visible for {\em 23EF} at 86 GHz in the edge on view for SgrA*-like system for an accretion rate which is slightly lower than those for {\em 20EF} and {\em 21EF}. for the same system at lower inclination angles and face-on views, the emission profile is broader and extends to much larger scales. For M87-like system too, with an accretion rate lower by an order than {\em 20EF} and {\em 21EF}, the emission profile is similar or slightly broader for {\em 23EF} at 230 GHz. Such a broader emission profiles for a similar physical set-up could indicate the presence of a disk wind which is due to velocity dispersion within the wind or jet.

In addition to tracing jets and wind signatures deduced from the width of the emission profiles, there are several interesting observations from these emission maps. As an example, the upper left panel of Fig.[\ref{fig:M87SgrAang230}] shows the face-on views of the emission where the unlensed photon ring and the outer and inner boundaries (corresponding to the prograde and retrograde spins) of the lensed photon ring are marked by green contours of radii $3~R_g$, $5.2~R_g$ and $6.2~R_g$, respectively. The lensed photon ring visible in face-on views ($0\degr$) coincides with the $5.2~R_g$ radius for almost all the cases (as expected for prograde spins).  The presence of a disk wind or jet base smear out the photon ring features as in {\em 21EF} for SgrA*-like system. As expected for face-on views, the flux profiles along different slices of the images coincide as can be seen in the lower left panel of Fig.[\ref{fig:M87SgrAang230}], while for small inclination angle such as $17^{\degr}$ the profiles separate and one of the peaks is higher than the other due to the additional effect of Doppler beaming. We also observe that for an inclination angle of $17\degr$, the emission peaks shifts from the expected inner or outer boundries of the lensed photon ring in all models except for {\em 22EF} which is an accreting Schwarzschild BH. The asymmetry is an intrinsic property of BHs with non-zero spins.

Simulations {\em 20}, {\em 21} and {\em 23} have very similar properties except for their floor density chosen (which are not expected to influence the emission signatures due to the assumed $\sigma_{cut}$). Simulations {\em 24EF} and {\em 26EF} are however set-up considering a larger mean magnetic field strength. Thus in general the accretion rates deduced for these simulation is much lower than those for the ones mentioned above as synchrotron emissivity is also dependent on the magnetic field strength. It is also observed that the accretion rates obtained from fitting the simulated spectrum to the data for M87 and SgrA* for {\em 24EF} and {\em 26EF} are of the same order for both cases but the emission region for {\em 26EF} is quite compact compared to {\em 24EF}. The magnetic field strength in case of {\em 26EF} is an order of magnitude higher, thus the emission from a smaller region is enough to obtain a similar amount of flux strength as obtained that from {\em 24EF}. All our simulations are set-up for Kerr BHs except for {\em 22EF} which is a spin 0 Schwarzchild BH. All other parameters for this simulation are same as those set up for simulations {\em 20}, {\em 21} and {\em 23}. Being a Schwarzschild BH, the ISCO radius is much larger implying that for similar accretion rates, the temperature of the in-falling material will be lesser than its Kerr counterpart. Thus to compensate for this in order to obtain similar fluxes, a higher accretion rate is required for both M87-like and SgrA* like systems.

Simulations {\em 21EF} and {\em 24EF} display very similar emission features at all frequencies, both in the edge-on views and low inclination angle views for both M87 and SgrA* like systems. {\em 24EF} although has a lower accretion rate compared to {\em 21EF} by more than an order of magnitude in both the sources, but {\em 24EF} also has an intrinsic magnetic field strength which is an order of magnitude higher. It seems that these two effects almost balance each other.

\begin{figure*}
 \begin{center}
 \includegraphics[height=3.0in,width=3.5in]{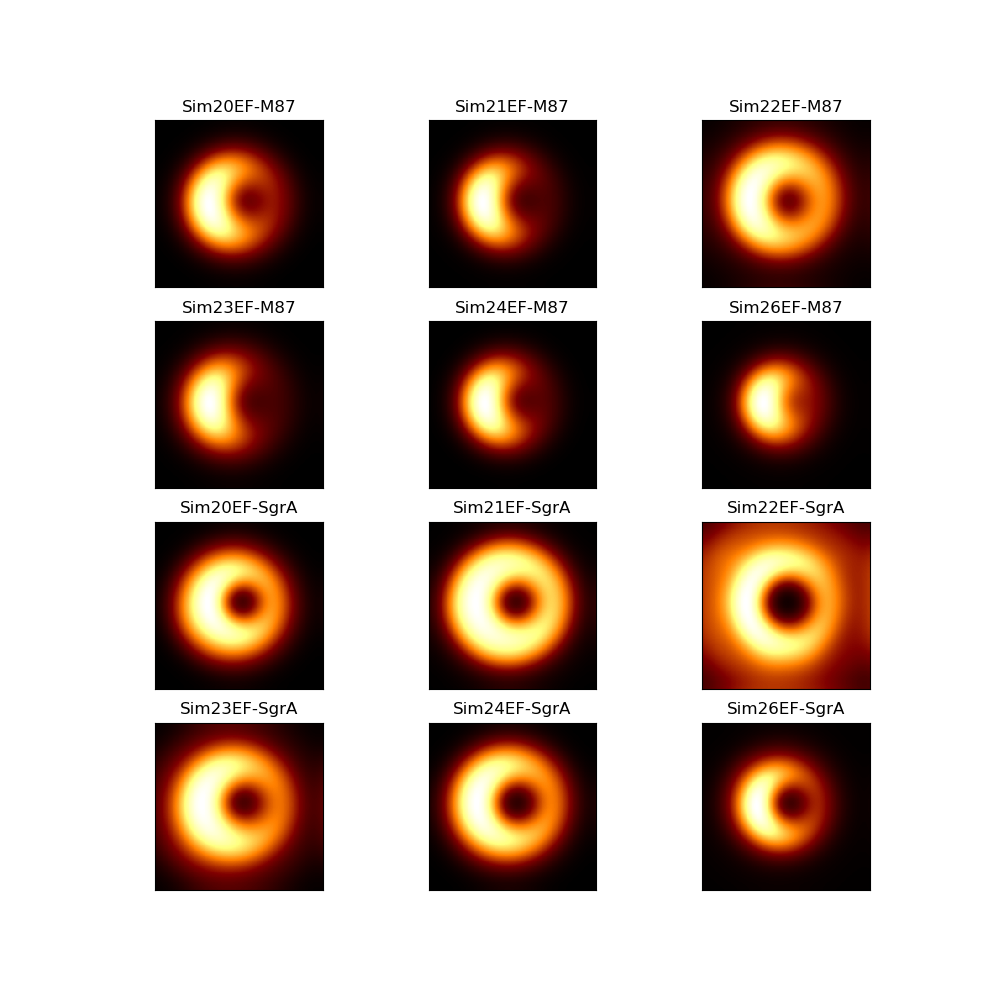}
 \includegraphics[height=3.0in,width=3.5in]{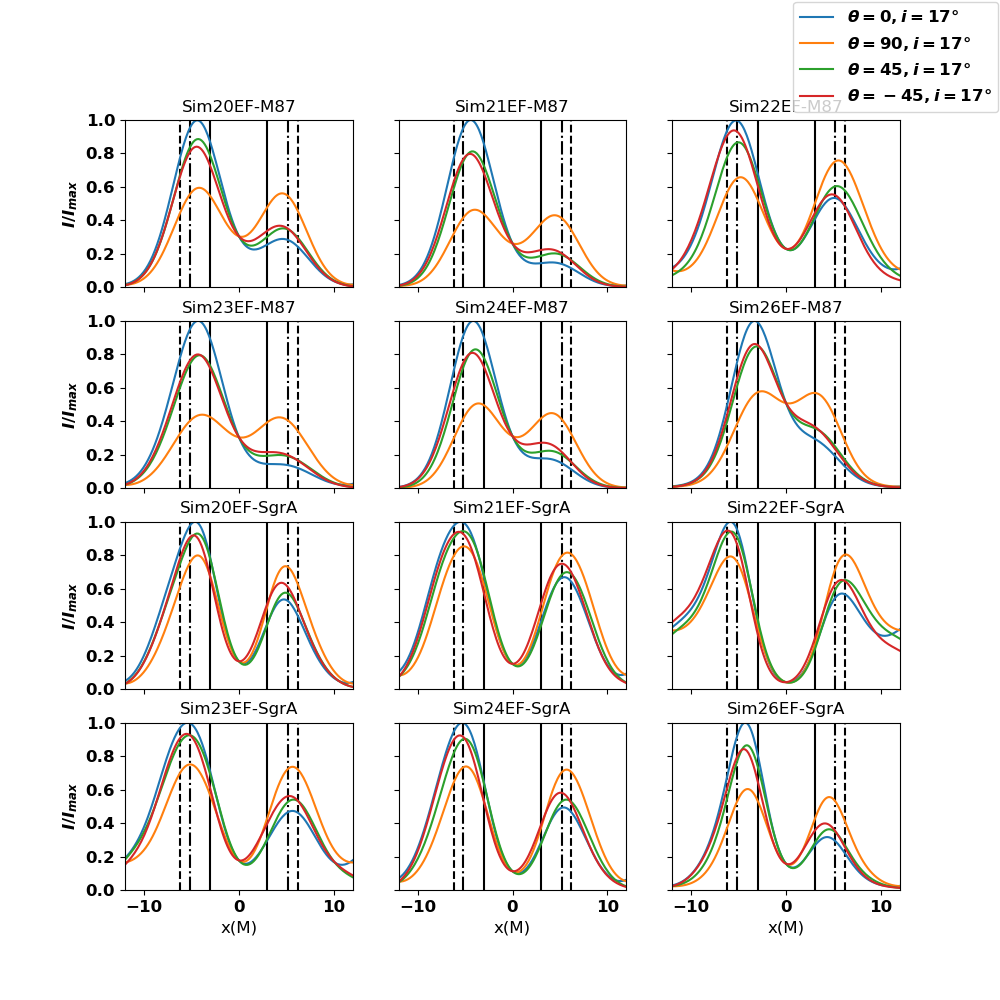} \\
 \includegraphics[height=3.0in,width=3.5in]{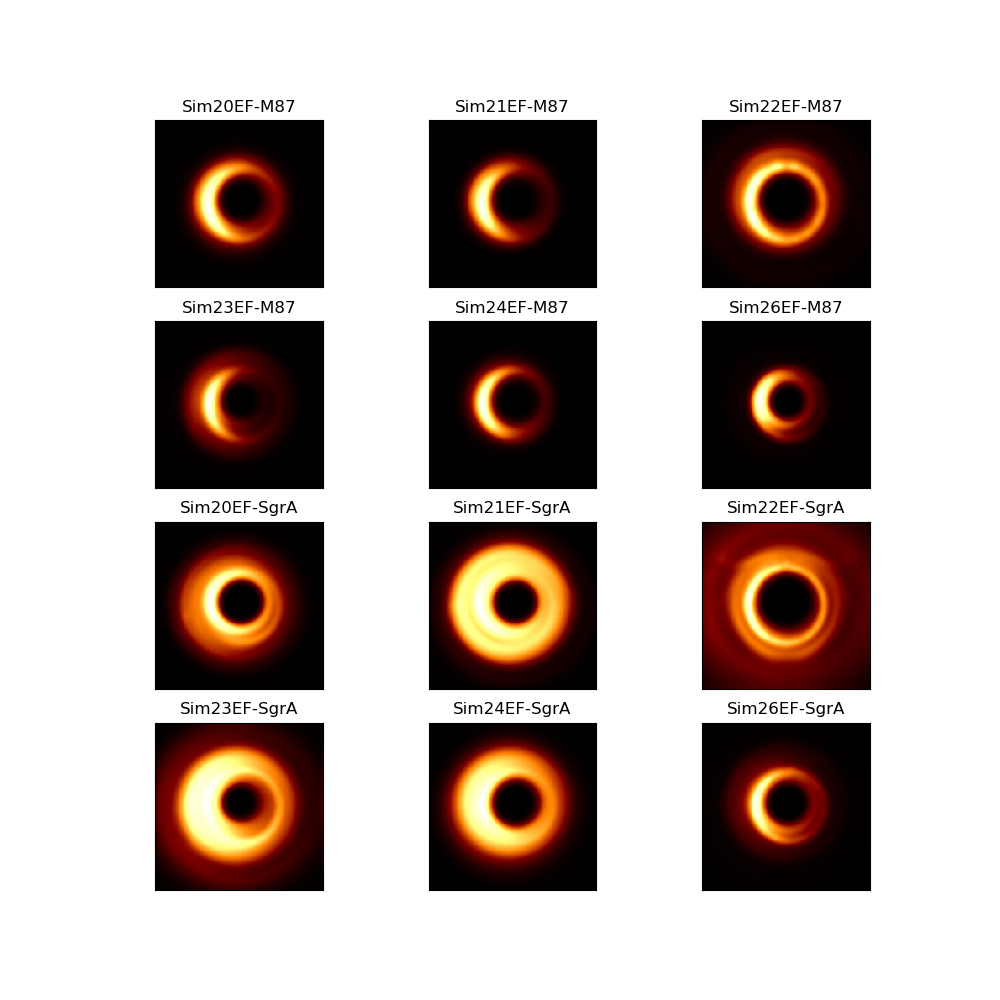} 
 \includegraphics[height=3.0in,width=3.5in]{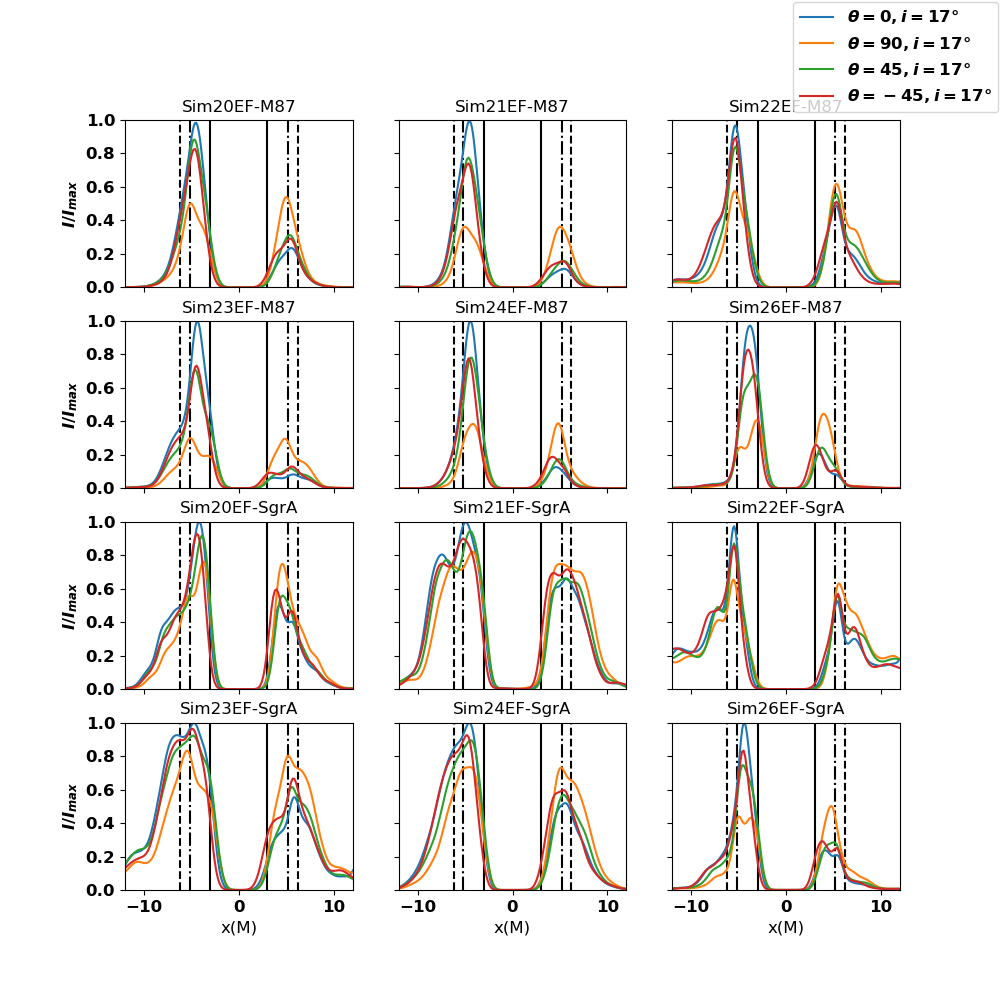}\\
 \includegraphics[height=3.0in,width=3.5in]{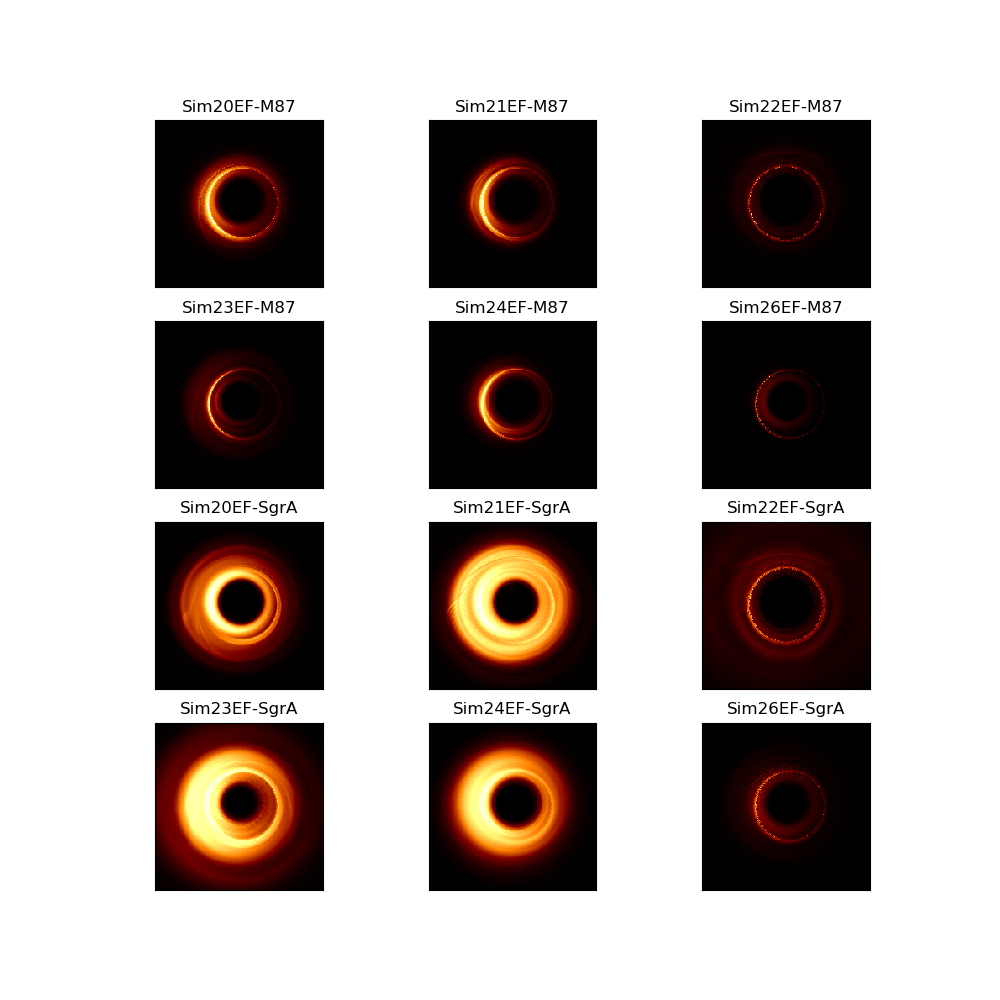}
 \includegraphics[height=3.0in,width=3.5in]{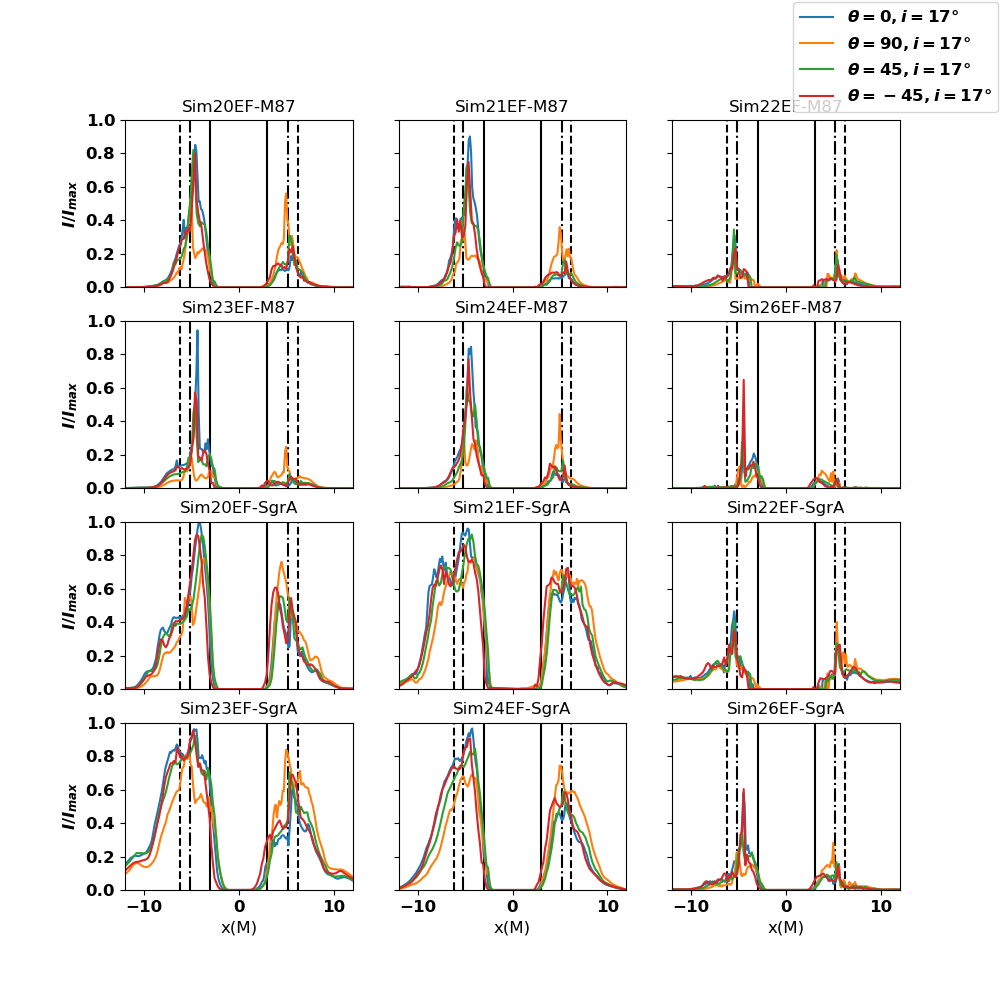}\\
  \caption{Emission maps and radial emission profiles (along diameters with angles$=0\degr, 90\degr, 45\degr \& -45\degr $ respectively) for M87 (upper 6 in each panel) and SgrA* (lower 6 in each panel) like systems for inclination angles of $17\degr$ at 230 GHz for EHT-$20~\mu$arcsec (upper), Geo-VLBI-$5~\mu$arcsec (middle) and L1-VLBI-$0.16~\mu$arcsec (lower) like resolutions). The intensity profiles in addition also mark the positions of the unlensed photon ring (black solid line) and the theoretical values of the inner (dot dashed line) and outer (dashed line) boundaries of the lensed photon ring.}\label{fig:M87SgrAL1VLBI}
\end{center}
\end{figure*}

The images obtained at lower inclination angles are further processed to account for the resolution of the current EHT($20~\mu \rm as$) and for the future earth-space based VLBI resolutions ($5~\mu \rm as$ for Geo-VLBI and $0.16~\mu \rm as$ for L1-VLBI) as shown in Fig.[\ref{fig:M87SgrAL1VLBI}], to predict if the origin of emission will be resolved today or in future for a better understanding of the physical processes. 

Given the current EHT resolution, {\em 20EF}, {\em 21EF}, {\em 23EF} and {\em 24EF} appear qualitatively indistinguishable (Fig.[\ref{fig:M87SgrAL1VLBI}]). {\em 22EF} shows large scale wings showing the signature from the large scale outflow while a much more compact emission profile is seen in {\em 26EF} for both M87-like and SgrA*-like systems. With the current EHT resolution, the shadow region is not well resolved for a M87 like system for {\em 23EF}, {\em 24EF} and especially {\em 26EF} whose emission region is quite compact. For the expected Geo-VLBI and L1-VLBI like resolutions, these similarities in the emission profiles seem to break down and and the intrinsic features start to appear (Fig.[\ref{fig:M87SgrAL1VLBI}]).

We would like to emphasize here that in this work we have considered only a thermal distribution for the electron distribution function for a simplistic approach. In reality we expect the jet emissions to be non-thermal in nature due to the presence of non-thermal processes such as magnetic field driven turbulence, magnetic reconnection, etc. We would however like to note that for simulations 20EF, 21EF, 22EF and 23EF this might be a good approximation which have a higher values of plasma $\beta$ ($\beta_0=P_{gas}/P_{mag}=10$) However, simulations 24EF and 26EF have $\beta=1$ and $\beta=0.1$, respectively, implying that the magnetic pressure is equal to or ten times greater than the gas pressure. In these it might not be sufficient to just consider a thermal distribution function for the electrons. For a true understanding of the origin of disk and wind signatures of a realistic system, it might not be sufficient to just consider the emissions originating from a population of thermal electrons. However, considering other distribution function is beyond the scope of this investigation.

\section{Comparison study}  \label{sec:Comp} 
The emission region for a M87-like system in general is more compact compared to SgrA* like systems in $R_g$ scales.
As can be seen in fig.[\ref{fig:M87SgrAedgeon}], {\em 22EF} shows a large scale outflow both for M87 and SgrA like systems visible prominently at 86 GHz. Similarly, {\em 23EF} shows a disk wind resulting in a large scale outflow visible distinctly at 86 GHz for both systems. {\em 21EF} displays a strong disk wind signature and a trace of the BZ jet along the upper half of the equatorial plane at 86 GHz (left panel of Fig.[\ref{fig:M87SgrAedgeon}]) in an edge-on view for SgrA* like system. The trace of the BZ jet is also visible at 230 GHz in the emission profile with the face-on view ($0\degr$) (bottom left panel of Fig.[\ref{fig:M87SgrAang230}]) where a non-zero flux is visible in the expected shadow region. {\em 20EF} and {\em 21EF} both have similar accretion rates but the emission profiles at 230 GHz display a broader emission profile of {\em 21EF} for SgrA like system and a greater Doppler beaming effect for M87 which leads to the conclusion that such a feature can arise if there is emission from region with additional velocities such as disk wind or outflows.

Similarly {\em 23EF} has a lower accretion rate than {\em 20EF} but broader emission profiles at 86 and 230 GHz for SgrA like systems and a comparable width in the emission profile for M87 at 86 GHz and 230 GHz. This is again possible if the emission has a contribution from the disk wind. At higher frequencies such as 700 GHz, however, we see contrary effects in the case of {\em 23EF} both for M87 and Sgr A like systems where the emission feature is peaked near the photon ring and is dimmer than {\em 20EF} or {\em 21EF} which is possible when the emission is from the accreting disk because at higher frequencies it is the energetics of the accreting disk which becomes relevant.

%

\section{Discussion and Conclusions}\label{sec:Conclu}
Most AGN are found to exist in nature with jets and outflows which in some cases are detected easily. It is thus important to understand the physical processes which launch different types of jets or outflows for which it is also important to trace their origin through observations. Through this work we have tried to demonstrate how a set of accreting systems with different physical conditions, results in varied observational features for systems with similar BH mass and spectral behaviours as M87 and SgrA*. BH shadows for both SgrA* and M87 have already been observed with the EHT \citep{EHTC2019a,EHTC2022a}. In this work our intention is not to make a quantitative comparison with the observations. However, this work intends to motivate how the same set of simulations, with similar spectral behaviour under different physical conditions, result in generating completely different observational signatures. In addition this work also intends to describe qualitative methods to extracting jet and wind signatures from observations. In our study the BH masses and SED are the primary constraining parameters. The accretion rates are varied in each simulation until the modeled SED is similar to the observed data. This results in generating intrinsic features in the emission maps, thus enabling to compare the different physical processes which lead to such emission signatures. The emission maps are further processed to make predictions for current and future VLBI observations. In general one of the criteria for accepting theoretical models is in comparing the jet power \citep{EHTC2019e} to the real values; however this investigation is more qualitative in its approach of demonstrating observational signatures of winds and jets, rather than testing alternate models for M87 and SgrA*. Thus in this work we do not calculate the jet power from these models as it is not relevant for the current investigation.

The emission maps here are shown in linear scale as generally done in the literature with the normalization being done with the maximum intensity in each frame. However we would like to mention that the mean value of intensity per pixel (which is quite low$~10^{-7}$ Jy) is similar for all the frequencies. A normalization with the mean value generates similar emission maps displaying a thick emission ring at all frequencies but the fine structures arising due to the presence of winds and jets are not visible at all frequencies. However, normalizing the images with the mean saturates the images and is not practical because the value is much less than the flux sensitivity of any telescope. On this note we would like to emphasize that even though we have taken into account the smudging of images due to the limited beam-width introduced by the resolution of telescopes, in reality the images obtained are further limited by thermal noise, limited $u-v$ coverage and calibration errors.

We would like to emphasize here that while post-processing, in addition to the accretion rate, it is also important to consider a proper electron distribution function depending on the underlying physical parameters of the GR-MHD which may affect the resulting spectrum significantly. Although considering a thermal electron distribution function may be sufficient for a weaker magnetic field limit but may not hold true in the strong field limit. In the strong field limit, we expect non-thermal processes to dominate arising due to effects such as magnetic re-connection and also holds true for the emission from jets where the magnetic fields dominate. In such cases considering a pure thermal electron distribution function will be an underestimation.

The outflow features are easily distinguishable at lower frequencies but tracing them to their origin from regions close to the BH would mean observing them with the highest resolutions which is possible only at higher frequencies and using telescopes with larger baselines. Currently highest resolution observations are possible with the technique of radio interferometery with different VLBI arrays such as the EHT which has an Earth sized baselines. Through this investigation we deduce that with fluxes like that we observe for M87 or SgrA*, observing the signatures of jets or outflows closest to horizon scale will be possible at 230 GHz. It might be possible to observe some of the features even at 345 GHz but currently not all EHT telescopes are designed to function at 345 GHz, which thus will introduce additional noise during image reconstruction. Although making strong deductions about different physical scenarios will not be possible with the current EHT resolution, it may however be possible with future Earth-Space baseline probes with ngEHT.

Apart from improving telescope resolution to obtain finer details of emissions from horizon scale around SMBHs with longer baselines and tuning the telescopes to observe at higher frequencies, it is also important to improve on the image reduction algorithms which generates the observed images with better resolution. In addition it is also important to consider the telescope sensitivity to obtain the threshold to distinguish the finer structures.

As a final note we would like to mention that through this work we primarily intended to demonstrate whether with current and future VLBI arrays it would be possible to detect signatures of disk winds or jets at horizon scales for two different physical systems. In the process we thus used a set of disk-wind-jet simulated models for AGNs to show how each model would differ from the other observationally. The SED was the primary constraint that was used and only the accretion rate was varied to fit the modeled SED to the observed spectrum. A more quantitative complete approach would include considering the variation in many other physical parameters (such as a considering different temperature models or distribution function for electrons etc.) to obtain precise fit to the SEDs. However for this investigation we think a qualitative approach is good enough to demonstrate the tools to extract certain observational features from simulated emission maps and making predictions if such an analysis would be possible for the current and future VLBI facilities.

\section*{Acknowledgements}

We acknowledge funding from ANID Chile via Nucleo Milenio TITANs (NCN19$-$058), Fondecyt 1221421, and Basal projects AFB-170002 and FB210003.

All GR-MHD simulations were performed on the VERA clusters of the Max Planck Institute for Astronomy. 
C.F. thanks Qian Qian and Christos Vourellis for developing the resistive version rHARM-3D of the original GR-MHD code HARM-3D, 
kindly provided by Scott Noble.


\section*{Data Availability}

 Data available on request. The data underlying this article will be shared on reasonable request to the corresponding author.



\bibliographystyle{mnras}
\bibliography{M87SgrA} 




\appendix

\section{Emission maps and profiles for 345 GHz and 700 GHz}
\begin{figure*}
 \begin{center}
 \includegraphics[height=3.5in,width=3.5in]{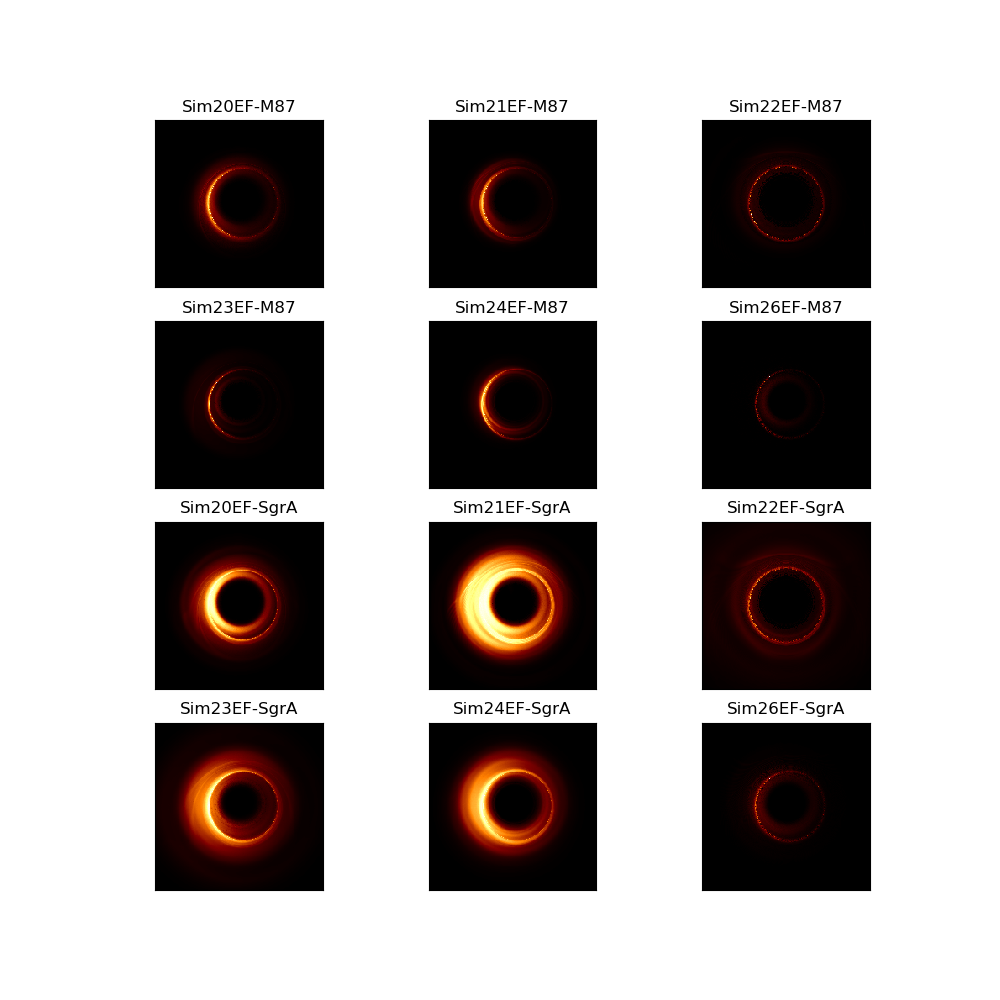}
 \includegraphics[height=3.5in,width=3.5in]{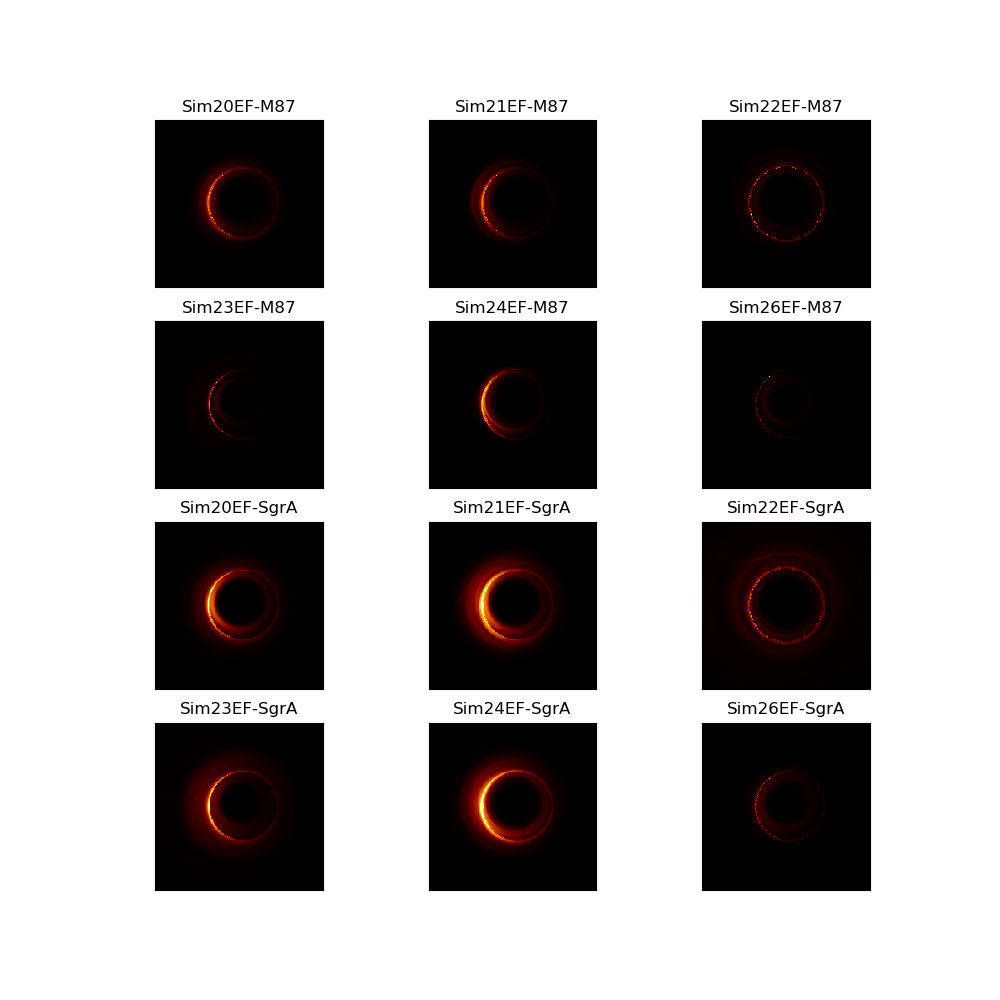}\\ 
 \includegraphics[height=3.5in,width=3.5in]{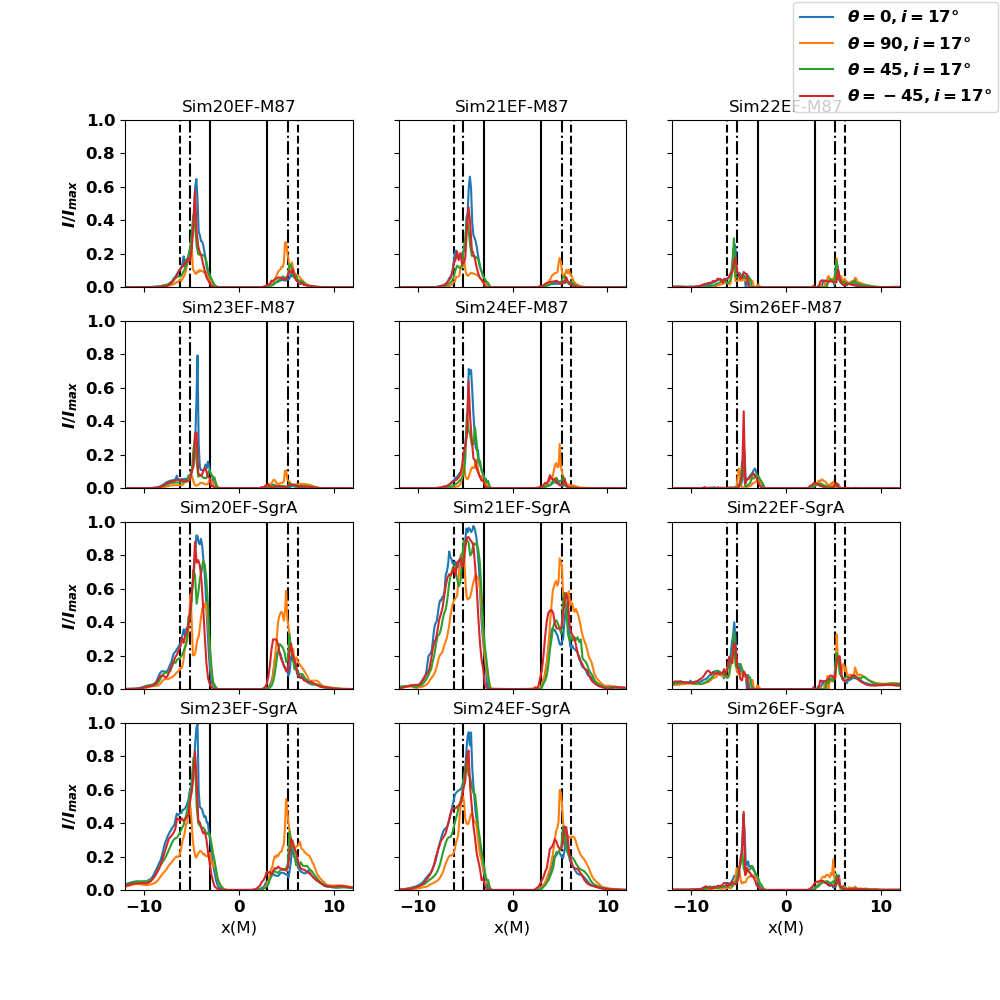} 
 \includegraphics[height=3.5in,width=3.5in]{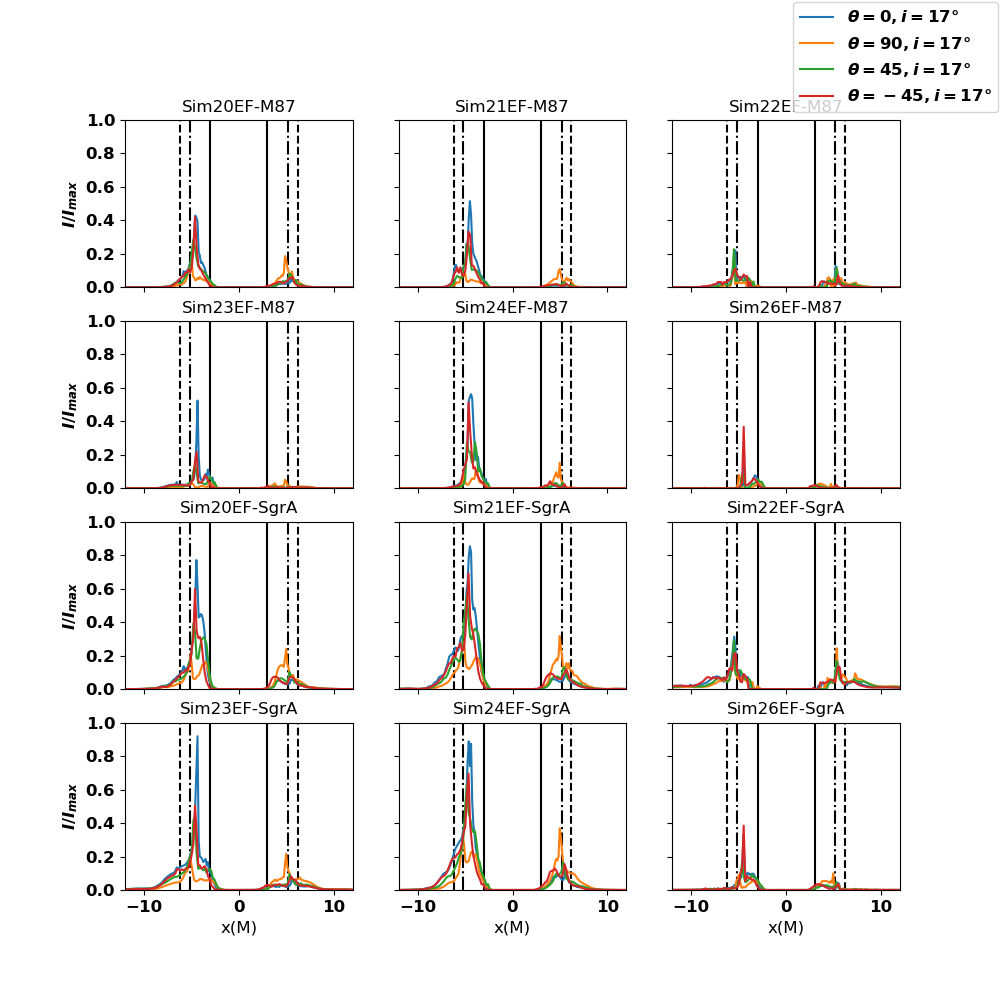}
  \caption{Flux normalized emission maps (upper) and intensity profiles (lower) (along diameters with angles$=0\degr, 90\degr, 45\degr \& -45\degr $ respectively) for all our simulations for M87 (upper 6 in each of the panels) and and SgrA* (lower 6 in each of the panels) like systems with inclination angle $17\degr$ for 345 GHz (left) and 700 GHz (right) respectively. The intensity profiles in addition also mark the positions of the unlensed photon ring (black solid line) and the theoretical values of the inner (dot dashed line) and outer (dashed line) boundaries of the lensed photon ring.} \label{fig:M87SgrAang345n700}
\end{center}
\end{figure*}

In Fig.[\ref{fig:M87SgrAang345n700}], we show the modeled emission maps and intensity profiles for M87-like and SgrA*-like systems from our simulations for 345 GHz and 700 GHz. The figure is discussed in Section [\ref{sec:Insten}].



\bsp	
\label{lastpage}
\end{document}